# Reversible Region of Non-Interest (RONI) Watermarking for Authentication of DICOM Images

*Jasni Mohamad Zain*[†] *and Malcolm Clarke*[††]

[†]*Faculty of Computer Systems and Software Engineering, Universiti Malaysia Pahang, Locked Bag 12, Kuantan, Pahang, Malaysia*
[††]*School of Information Systems, Computing and Mathematics, Brunel University, Kingston Lane, Uxbridge, Middlesex, UB8 3PH, United Kingdom.*

**Summary**
This paper introduces current watermarking techniques available from the literatures. Requirements for medical watermarking will be discussed. We then propose a watermarking scheme that can recover the original image from the watermarked one. The purpose is to verify the integrity and authenticity of DICOM images. We used ultrasound (US) images in our experiment. SHA-256 of the whole image is embedded in the least significant bits of the RONI (Region of Non-Interest). If the image has not been altered, the watermark will be extracted and the original image will be recovered. SHA-256 of the recovered image will be compared with the extracted watermark for authentication.
*Key words:*
*DICOM, image authenticity and integrity, watermarking.*

## 1. Introduction

Modern health care infrastructure is based on digital information management. The digital imaging and communication in medicine (DICOM) standard facilitates the communication of digital image information regardless of device manufacturer. It is usual that a medical image is diagnosed before storing the image in the long-term storage, so the significant part of the image is already determined [1]. The significant part is called ROI (Region of Interest). Although the recent advancement in information and communication technologies provide new means to access, handle and move medical images, they also allow easy manipulation and replication [2]. It is common view that there is an urgent need of security measures in medical information system.

Digital watermarking can imperceptibly embeds messages without changing image size or format. When applied for medical images, the watermarked image can still conform to the DICOM format [3]. Some researchers already apply watermarking technique for medical data. Zhou et al[4] present a watermarking method for verifying authenticity and integrity of digital mammography image. They used digital envelope as watermark and the least significant bits (LSB) of one random pixel of the mammogram is replaced by one bit of the digital envelope bit stream. Instead of the whole image data, only partial image data, i.e. the most significant bits (MSB) of each pixel is used for verifying integrity. Other researchers adapt digital watermarking for interleaving patient information with medical images to reduce storage and transmission overheads [5]. Again, the LSB of image pixels are replaced for embedding. Chao et al. propose a discrete cosine transform (DCT) based data-hiding technique that is capable of hiding those EPR related data into a marked image [6]. The information is embedded in the quantized DCT coefficients. The drawback of the above watermarking approaches is that the original medical image is distorted in a non-invertible manner. Therefore it is impossible for watermark decoder to recover the original image.

A reversible watermarking scheme involves inserting a watermark into the original image in an invertible manner in that when the watermark is extracted, the original image can be recovered completely [7][8][9][10]. Research has also been done in the area of reversible watermarking in medical images. Trichili et al[11] proposes an image virtual border as the watermarking area. Patient data is then embedded in the LSBs of the border. Guo and Zhuang present a scheme where the digital signature of the whole image and patient information is embedded [3]. Cao et al extend their work on digital envelope and embed their DE by making a random walk sequence and replace LSB of each selected pixel [12].

In this paper, we give an overview of watermarking techniques and discuss the requirements for medical watermarking. We then propose a lossless watermarking scheme being capable of verifying authenticity and integrity of DICOM images. Besides that the original image can be exactly recovered at the receiver site, the whole image's integrity can be strictly verified. In Section 2, we introduce watermarking techniques, followed by requirements for medical image watermarking in Section 3. In Section 4, we present our watermarking scheme, including data embedding, extracting and verifying procedure. In Section 5, experimental results are provided to demonstrate that such scheme can embed large payload while keeping distortion level very low.

## 2. Overview of Watermarking Techniques

Current watermarking techniques described in the literature can be grouped into three main classes. The first includes the spatial domain methods, which embed the watermark by directly modifying the pixel values of the original image. The second class includes the transform domain methods, which embed the data by modulating the transform domain signal coefficients. The transform domain techniques have been found to have the greater robustness, when the watermarked signals are tested after having been subjected to common signal processing. The third class is the feature domain technique. This technique takes into account region, boundary and object characteristics. Such watermarking technique may give additional advantages in terms of detection and recovery from geometric attacks, compared to previous approaches.

2.1 Spatial Domain Technique

The most straightforward method of watermark embedding would be to embed the watermark into the least significant bits of the cover object [13]. Given the extraordinarily high channel capacity of using the entire cover for transmission in this method, a smaller object may be embedded multiple times. Even if most of these were lost due to attacks, a single surviving watermark would be considered a success. LSB substitution however despite its simplicity brings a host of drawbacks. Although it may survive transformations such as cropping, any addition of noise or lossy compression is likely to defeat the watermark. An even better attack would be to simply set the LSB bits of each pixel to one fully defeating the watermark with negligible impact on the cover object. Furthermore, once the algorithm is discovered, an intermediate party could easily modify the embedded watermark. LSB modification proves to be a simple and fairly powerful tool, however lacks the basic robustness that watermarking applications require.

Another technique for watermark embedding is to exploit the correlation properties of additive pseudo-random noise patterns as applied to an image [14]. A pseudo-random noise (PN) pattern $W(x, y)$ is added to the cover image $I(x, y)$, according to the equation 1.

$$Iw(x,y) = I(x,y) + k* W(x,y) \qquad (1)$$

In equation.1, k denotes a gain factor, and IW the resulting watermarked image. Increasing k increases the robustness of the watermark at the expense of the quality of the watermarked image. Rather than determining the values of the watermark from "blocks" in the spatial domain, we can employ CDMA spread-spectrum techniques to scatter each of the bits randomly throughout the cover image, increasing capacity and improving resistance to cropping [14]. To detect the watermark, each seed is used to generate its PN sequence, which is then correlated with the entire image. If the correlation is high, that bit in the watermark is set to "1", otherwise a "0". The process is then repeated for all the values of the watermark. CDMA improves on the robustness of the watermark significantly, but requires several orders more of calculation.

2.2 Frequency Domain Techniques

The classic and still most popular domain for image processing is that of the Discrete-Cosine-Transform, or DCT. The DCT allows an image to be broken up into different frequency bands, making it much easier to embed watermarking information into the middle frequency bands of an image. The middle frequency bands are chosen such that they have minimize they avoid the most visual important parts of the image (low frequencies) without over-exposing themselves to removal through compression and noise attacks (high frequencies). One such technique utilizes the comparison of middle-band DCT coefficients to encode a single bit into a DCT block. To begin, we define the middle-band frequencies (FM) of an 8x8 DCT block as shown below in Fig. 1.

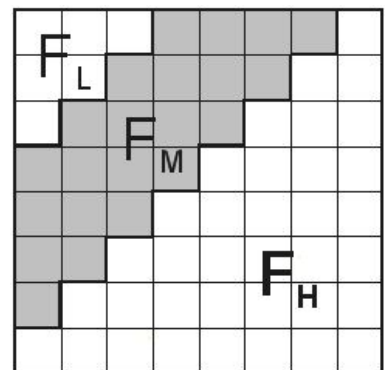

Fig. 1 Definition of DCT Regions

FL is used to denote the lowest frequency components of the block, while FH is used to denote the higher frequency components. FM is chosen as the embedding region as to provide additional resistance to lossy compression techniques, while avoiding significant modification of the cover image [15].

Another possible technique for watermark embedding is that of the wavelet transform. The DWT (Discrete Wavelet Transform) separates an image into a lower resolution approximation image (LL) as well as horizontal (HL), vertical (LH) and diagonal (HH) detail components. The process can then be repeated to computes multiple "scale" wavelet decomposition, as in the 2-scale wavelet transform shown below in Fig. 2.

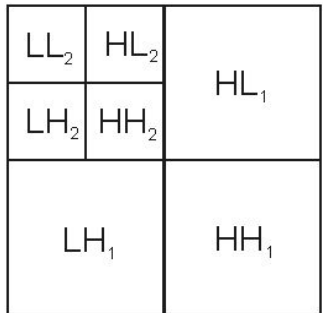

Fig. 2. 2-Scale Discrete Wavelet Transform

One of the many advantages over the wavelet transform is that it is believed to more accurately model aspects of the HVS as compared to the FFT or DCT. This allows us to use higher energy watermarks in regions that the HVS is known to be less sensitive to, such as the high resolution detail bands {LH,HL,HH). Embedding watermarks in these regions allow us to increase the robustness of our watermark [14].

One of the most straightforward techniques is to use a similar embedding technique to that used in the DCT, the embedding of a CDMA sequence in the detail bands according to equation 2.

$$I_{Wu,v} = \begin{cases} W_i + \alpha |W_i| x_i, & u,v \in HL, LH \\ W_i & u,v \in LL, HH \end{cases} \quad (2)$$

Where $W_i$ denotes the coefficient of the transformed image, $x_i$ the bit of the watermark to be embedded, and a scaling factor. To detect the watermark we generate the same pseudo-random sequence used in CDMA generation and determine its correlation with the two transformed detail bands. If the correlation exceeds some threshold T, the watermark is detected.

This can be easily extended to multiple bit messages by embedding multiple watermarks into the image. As in the spatial version, a separate seed is used for each PN sequence, which is then added to the detail coefficients as in equation 2. During detection, if the correlation exceeds T for a particular sequence a "1" is recovered; otherwise a zero. The recovery process then iterates through the entire PN sequence until all the bits of the watermark have been recovered. Furthermore, as the embedding uses the values of the transformed value in embedded, the embedding process should be rather adaptive; storing the majority of the watermark in the larger coefficients. The technique should prove resistant to JPEG compression, cropping, and other typical attacks.

2.3 Feature Domain Techniques

The two other techniques have been mainly focused on applying the watermarking on the entire image domain. This technique was developed in order to increase robustness. The method takes into account region, boundary and object characteristics and gives additional advantages in terms of detection and recovery from geometric attacks compared to previous methods.

Kutter et al. [16] used feature point extraction and the Voronoi diagram as an example to define region of interest (ROI) to be watermarked. The feature extraction process is based on a decomposition of the image using Mexican-Hat mother wavelet, as shown in Fig. 3.

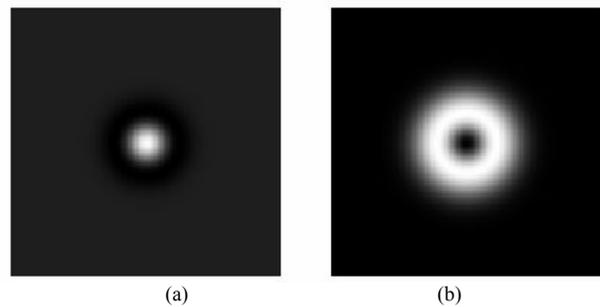

(a)               (b)
Fig. 3. Mexican-hat mother wavelet function in spatial domain (a) and in transform domain (b)

In two dimensions the Mexican-hat wavelet, $\psi(\varpi)$, at location $(\varpi)$, is defined as:

$$\psi(\varpi) = (2 - |\varpi|^2) \exp(-\varpi^2/2) \quad (3)$$

where $\varpi$ is the two-dimensional coordinate of a pixel. Then the wavelet in the spatial-frequency domain can be written as:

$$\psi_H(\check{n}) = (\check{n} \cdot \check{n}) e^{-1/2(\check{n} \cdot \check{n})} \quad (4)$$

where $\check{n}$ is the 2D spatial-frequency variable. The Mexican Hat is always centered at the origin in the frequency domain, which means that the response of a Mexican Hat wavelet is invariant to rotation. However, the stability of the method proposed depends on the feature points. These extracted features have drawback that their location may change by some pixels because of attack or during the watermarking process. Changing the location of the extracted feature points will cause problems during the detecting process.

## 3. Requirements for Medical Image Watermarking

Security of medical images, derived from strict ethics and legislative rules, gives rights to the patient and duties to the health professionals. This imposes three mandatory characteristics: confidentiality, reliability and availability:
• Confidentiality means that only the entitled persons have access to the images;
• Reliability which has two aspects; Integrity: the image has not been modified by non-authorized person, and authentication: a proof that the image belongs indeed to the correct patient and is issued from the correct source;
• Availability is the ability of an image to be used by the entitled persons in the normal conditions of access and exercise.

Security risks of medical images can vary from random errors occurring during transmission to lost or overwritten segments in the network during exchanges in the intra- and inter-hospital networks. One must also guarantee that the header of the image file always matches that of the image data. In addition to these unintentional modifications one can envision various malicious manipulations to replace or modify parts of the image, called tampering. The usual characteristics of watermarking are invisibility of the mark, survives common distortions, carries many bits of information, secrecy to unauthorized persons, and requires little computation to insert or detect [17]. These demands also exist in the medical domain but additional constraints are added. Three main objectives are foreseen in the medical domain [18][19]:

### 3.1 Imperceptible / Reversible Watermarking

Medical tradition is very strict with the quality of biomedical images. Thus the watermarking method must be reversible, in that the original pixel values must be exactly recovered [20]. This limits significantly the capacity and the number of possible methods. An alternative way is to define regions of interest, to be left intact, and leave us with regions of insertion where watermark could be inserted and does not interfere or disturb the radiologist.

### 3.2. Integrity Control

The "trustworthy camera" concept applies also to medical images, especially in the context of legal aspects and insurance claims. Friedman has used this concept in his work [21]. By embedding an encryption chip in the camera, the camera endorses its captured pictures and generates content-dependant digital signatures. There is thus a need to prove that, the images on which the diagnoses and any insurance claims are based have preserved their integrity.

### 3.3. Authentication

A critical requirement in patient records is to authenticate the different parts of the electronic patient record, in particular the images. More often an attached file or a header, which carries all the needed information, identifies an image. However, keeping the meta-data of the image in a separate header file is prone to forgeries or clumsy practices. An alternative would be to embed all such information into the image data itself.

## 4. Methodology

### 4.1 Watermark

The watermark is generated by creating a hash value from the region of interest (inside the rectangle), X of size m x n. The pixels will be arranged in a string, S.

$$S = B(X_{(1,1)} X_{(1,2)} ... X_{(1,m)} X_{(2,1)} ... X_{(m,n)}) \quad (5)$$

where Xmn is the 8 bit binary value of each pixel.

The hash value is obtained by applying a hash function to the string

$$Hash = H(S) \quad (6)$$

where H is any hash function such as MD5 and SHA256.

### 4.2 Embedding Region and Domain

The embedding region is considered to be outside the region of interest in order to prevent distortion to the area as a result of adding the watermark. In an ultrasound image, the embedding region is normally a dark region with pixel values 0. This feature will be exploited to create a reversible or invertible watermarking.

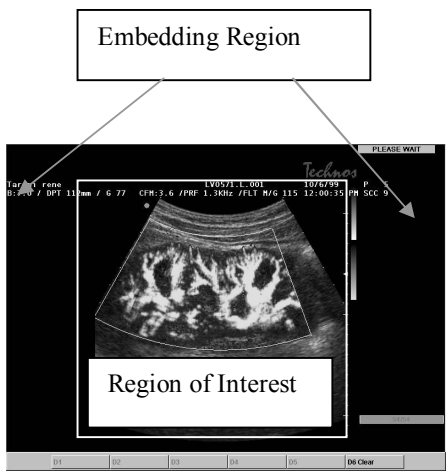

Fig. 4. Embedding region

In strict authentication watermarking, it is vital that the system will detect any change to the image. Fragile watermarking is the most appropriate as any change in the image will also affect the watermark. Least Significant Bit (LSB) watermarking has an advantage as the method of choice, as it is well known that LSB is vulnerable and easy to manipulate.

### 4.3 Security

A watermark is secure if it is able to resist intentional tampering by an attacker. This would include remaining secure even when the attacker knows the algorithm for embedding and extracting the watermark.

The strength of the security of the watermark will depend on the key chosen. A typical attack would involve removing the watermark, changing the image, then recalculating and embedding the new hash value into the embedding area. If the key for calculating the hash value remains secret, then the system may be considered secure. The secret key can be used to create the hash value and to create a random embedding. These will be examined in turn.

- Key for hashing

A key can be used to create the hash for the selected region. In this method, the sender and recipient will use the same key to carry out the hash function. The hash value obtained will be used as the watermark. At the recipient end, the key will be used to carry out the hash function on the received image and the hash value will be compared with the hash extracted.

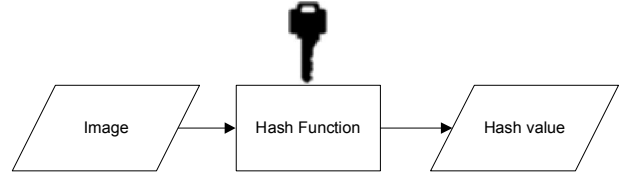

Fig. 5. Key for hash

- Key for embedding

A key used for embedding will determine the random mapping of watermark values into the embedding region as in fig. 6.

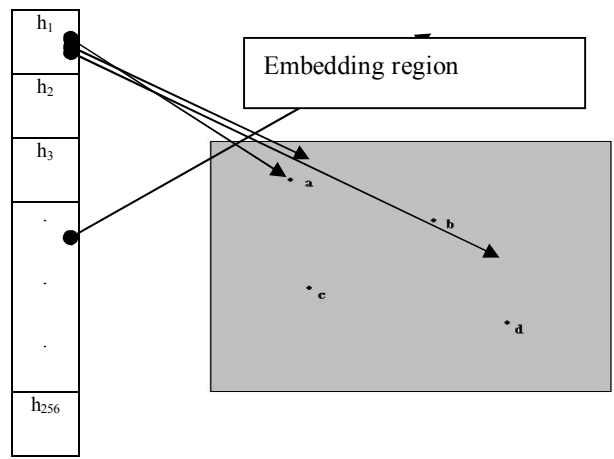

Fig. 6. Hash value mapping in the embedding region

This supposes that the number of points or pixels in the embedding region is greater than or equal to the number of bits in the hash value holds. As an example, suppose the pixels are arranged as a simple raster scan as in fig. 7.

| 1 | 2 | 3 | 4 | 5 |
|---|---|---|---|---|
| 6 | 7 | 8 | 9 | 10 |
| 11 | 12 | 13 | 14 | 15 |
| 16 | 17 | 18 | 19 | 20 |

Fig. 7. Embedding region of 5 x 4 pixels

which may be described by the mapping function of equation (7):

$$f(x) = x \mod n \qquad (7)$$

where x is the bit position and $x \in \{1, h\}$ and n is the total number of pixels available for embedding. In this example, we use h=20 to make full use of the embedding region. Applying equation 7, bit position one will be located in pixel number one, bit position two will be located in pixel number 2 and so on. By using a key, k, the position will be randomised. If a simple function, e.g. equation (8) is applied,

$$f(x) = kx \mod n \qquad (8)$$

where k is a prime key, then the mapping will be a randomised one-to-one mapping.

If the embedding region is 10 x 10 pixels, then the distribution of embedding will be pictured as in Fig. 8.

|    |    |    | 19 |    |    | 11 |    |    |    |
|----|----|----|----|----|----|----|----|----|----|
|    | 3  |    |    |    |    |    | 14 |    |    |
|    |    | 6  |    |    |    |    |    |    | 17 |
|    |    |    | 9  |    |    | 1  |    |    |    |
| 20 |    |    | 12 |    |    |    | 4  |    |    |
|    |    |    |    | 15 |    |    |    |    | 7  |
|    |    |    |    |    | 18 |    |    |    |    |
| 10 |    |    | 2  |    |    |    |    |    |    |
|    | 13 |    |    | 5  |    |    |    |    |    |
|    |    | 16 |    |    | 8  |    |    |    |    |

Fig. 8. Distribution of embedding for k=37, h=20, n=100

This simple method relies on the use of symmetric keys, which has an associated problem of key management. This is beyond the scope of this research. In practice asymmetric key systems are favoured; these are discussed in the next section.

### 4.4 Hashing – SHA256

The Secure hash Algorithm (SHA) was developed by the National Institute of Standards and Technology (NIST) and published as a federal information processing standard (FIPS PUB 180) in 1990. The algorithm is an iterative, one-way hash function that can process a message to produce a condensed representation called a message digest. The algorithm enables the integrity of a message to be determined and any change to the message will, with a very high probability result in a different message digest. This property is useful in the generation and verification of digital signatures and message authentication codes. It is based on a public/private key, and thus overcomes the problem of key management.

### 4.5 Algorithm

SHA-256 may be incorporated into a watermarking algorithm as shown in Figure 9. The general methodology and principles as listed below:

At sender site

1) Define Area: The Region of Interest (ROI) is determined as the smallest rectangle that bounds the known image area. Figure 4.1 shows an example of a rectangle defining the ROI in an ultrasound image.
2) SHA-256: The hash value for the whole image using SHA-256 is calculated. This produces a 256-bit one-way hash value that can be the basis of the watermark.
3) Embed: The hash value is embedded into the Region of Non-Interest (RONI) in the LSB. The specific location is not important, as it is known that it will not affect the image under any circumstances.

At receiver site:

1) Extract watermark: The watermark is extracted by recovering the LSB from the watermarking area.
2) Flipping: In flipping, the LSB in the watermarking area are reset to their original values. This acts as the reversible function and is possible for any image that has an area of known constant. In the case of ultrasound images, this may be easily achieved by resetting all the bits to zero.
3) SHA-256: the SHA-256 algorithm is applied to the received image and the hash value computed.
4) Authentication: the hash value calculated in step 3 is compared to that extraction in step 2. If found to be the same, the image is authenticated.

## 5. Experimental Results

An 800 x 600 pixels ultrasound image was watermarked using the method described in section 4. The watermarked image was modified using the cloning tools of Adobe Photoshop CS2. The cloning area was around 50x50 pixels and the change may be seen as the image in Fig. 9. Fig. 10 shows the results of hashing using SHA-256.

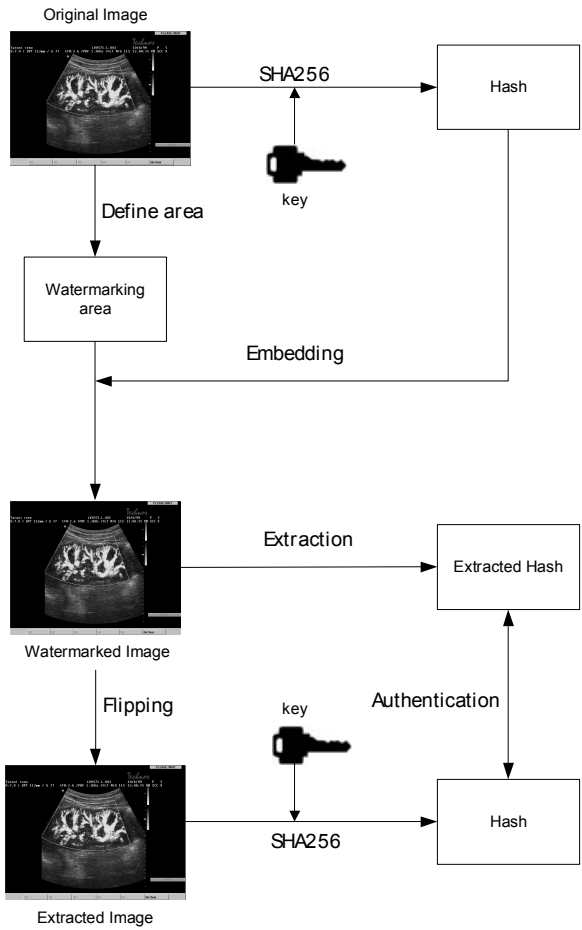

Fig. 9. Strict Authentication Watermarking (SAW) System

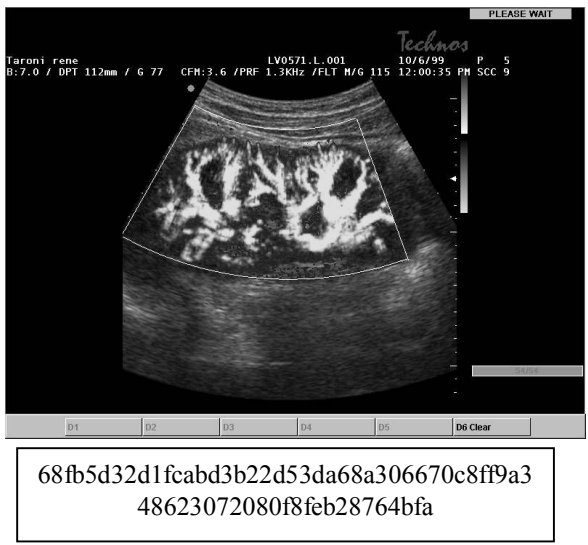

68fb5d32d1fcabd3b22d53da68a306670c8ff9a3
48623072080f8feb28764bfa

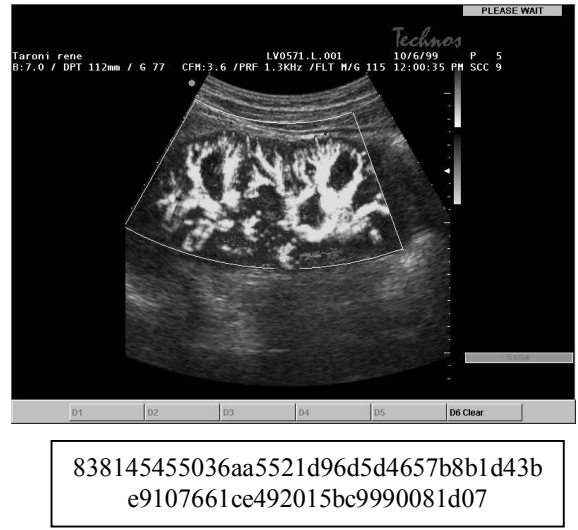

838145455036aa5521d96d5d4657b8b1d43b
e9107661ce492015bc9990081d07

Fig 10 (a) Original image and its hash (b) Tampered image and its hash

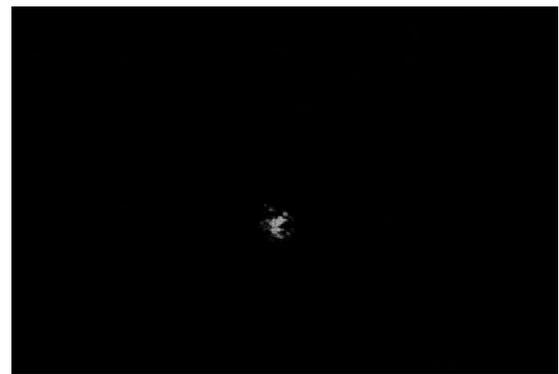

Fig. 11 Image difference

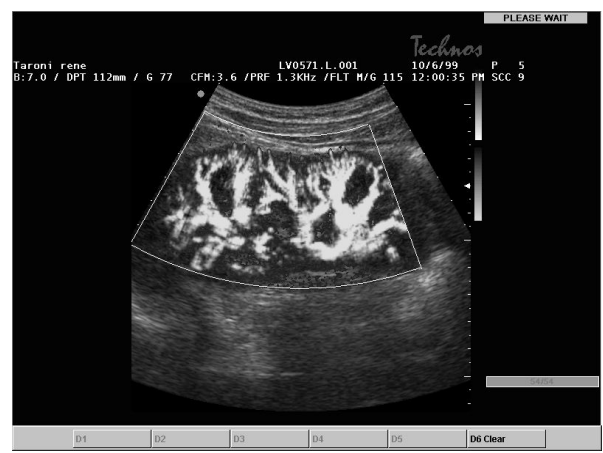

Fig. 12 Watermarked image with 550kb payload

Two blocks were then watermarked, using one LSB and two LSBs, increasing in the number of bits embedded to determine the capacity of LSB embedding before the recommended PSNR of 32dB was reached. Table 1 shows the result of embedding 270kb up to 550kb in the region of non-interest.

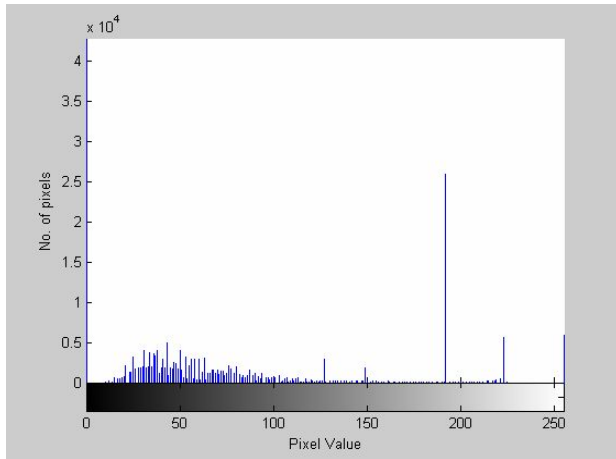

Fig.13(a) Histogram of Original Image

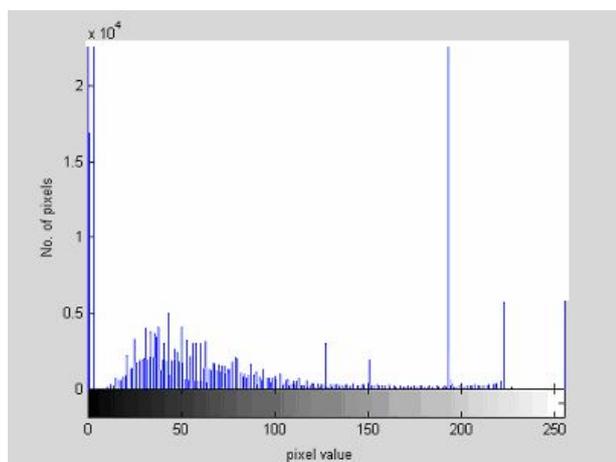

Fig.13(b) Histogram of watermarked image (550kb)

Fig. 13 (a-b) shows the image histogram of the original image and a watermarked image with 550 kb payload. The histogram clearly shows the dramatic increase in the pixels with values 1 and 3, but keeping the remaining pixels exactly the same.

Table 1 Capacity and PSNR for 800x600 US Image

| Capacity (kb) | PSNR (dB) |
|---|---|
| 270 | 249.6 |
| 430 | 51.5 |
| 475 | 42.9 |
| 510 | 31.7 |
| 550 | 27.4 |

## 6.0 Conclusion

Watermarking in medical images has a lot of potential. From the large capacity available for embedding, a lot more information can be added to the image to make it more secure. Few researchers have included patient's data and digital signature as watermark [4][11][12]. Combining cryptography and compression will add security and more information to the limited capacity. The most important thing for medical image communications is that the image is still conforms to the DICOM image format after watermarking takes place.

We proposed a lossless watermarking scheme being capable of verifying authenticity and integrity of DICOM images. Besides that the original image can be exactly recovered at the receiver site, the whole image's integrity can be strictly verified. We presented our watermarking scheme, including data embedding, extracting and verifying procedure. Experimental results showed that such scheme could embed large payload while keeping distortion level very low.

## References

[1] Wakatani, A. 2002, "Digital Watermarking for ROI Medical Images by Using Compressed Signature Image", 35th Annual Hawaii International Conference on System Sciences (HICSS-35'02), pp. 2043-2048.
[2] Coatrieux, G., Sankur, B. & Maitre, H. 2001, "Strict Integrity Control of Biomedical Images", SPIE Conf. 4314: Security and Watermarking of Multimedia Contents III.
[3] Guo, X. & Zhuang, T. 2003, "A lossless watermarking scheme for enhancing security of medical data in PACS", Medical Imaging 2003: PACS and Integrated Medical Information Systems: Design and Evaluation, Feb 18-20 2003, The International Society for Optical Engineering, San Diego, CA, United States, pp. 350-359.
[4] Zhou, X.Q., Huang, H.K. & Lou, S.L. 2001, "Authenticity and integrity of digital mammography images", IEEE Transactions on Medical Imaging, vol. 20, no. 8, pp. 784-791.
[5] Acharya, R., Anand, D., Bhat, S. & Niranjan, U.C. 2001, "Compact storage of medical images with patient information", IEEE transactions on information technology in biomedicine : a publication of the IEEE Engineering in Medicine and Biology Society, vol. 5, no. 4, pp. 320-323.
[6] Chao, H.M., Hsu, C.M. & Miaou, S.G. 2002, "A data-hiding technique with authentication, integration, and confidentiality for electronic patients records", IEEE Transactions Information Technology In Biomedicine, vol. 6, no. 1, pp. 46-53.

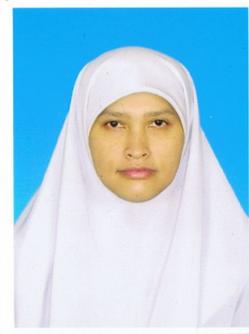

Jasni Mohamad Zain received her Bachelor degree in Computer Science from University of Liverpool, England, UK in 1989; PGCE Mathematics from Sheffield Hallam University, England, UK in 1994; M.E. degree from Hull University, England, UK in 1998 and PhD from Brunel University, West London, UK in 2005. She currently holds the post as the Director of the Center of Information Technology and Communication, University Malaysia Pahang. She is currently a lecturer in Faculty of Computer Science and Software Engineering, University Malaysia Pahang. She has been actively presenting papers in national and international conferences. Her research interests include Image Processing as well as Data and Network Security

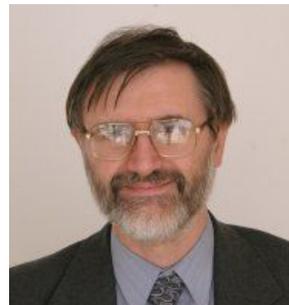

Malcolm Clarke is a Senior Lecturer in telemedicine and eHealth Systems in the School of Information Systems, Computing and Mathematics, Brunel University. He gained his PhD in medical engineering at Imperial College in 1984, developing and using a computerised 40 lead ECG acquisition system for total body surface potential mapping in ECG stress exercise testing. He then developed an ultrasound system for intra-arterial scanning. He moved to Brunel University in 1989 where he developed and led a Master's programme for data communications until 1999. Dr Clarke has a unique combination of expertise in communications, engineering and systems design with experience of working in the medical field for 20 years.